\documentstyle[prb,aps,multicol,epsf]{revtex}
\tighten
\begin{document}
\draft

\title{Shot-noise suppression by Fermi and Coulomb correlations in ballistic 
conductors}

\author{O. M. Bulashenko and J. M. Rub\'{\i}}
\address{Departament de F\'{\i}sica Fonamental,
Universitat de Barcelona, Diagonal 647, E-08028 Barcelona, Spain}

\date{April 6, 2001}
\maketitle

\begin{abstract}
We investigate the injection of degenerate Fermi-Dirac electrons
into a multimode ballistic conductor under the space-charge-limited regime.
The nonequilibrium current fluctuations were found to be 
suppressed by both Coulomb and Fermi correlations.
We show that the Fermi shot-noise-suppression factor is limited below by 
the value $2 k_BT/\varepsilon_F$, where $T$ is the temperature
and $\varepsilon_F$ the Fermi energy of the injected electrons.
The Coulomb noise suppression factor may attain much lower values
$\varepsilon_F/2qU$, because of its dependence on the applied bias 
$U\gg k_BT/q$.
The asymptotic behavior of the overall shot-noise-suppression factor
in a high degenerate limit was found to be $k_BT/qU$, 
independent of the material parameters.
\end{abstract}


\begin{multicols}{2}

\section{Introduction}

Nonequilibrium fluctuations of the electric current (shot noise) 
in mesoscopic conductors have received recently significant attention.
\cite{dejong97,blanter00} In particular, the shot noise in scattering-free 
or {\em ballistic} conductors has been studied extensively 
both theoretically \cite{lesovik89,buttiker9092,land-martin9192}
and experimentally, \cite{li90,reznikov95,kumar96,brom99}
by focusing mainly on the suppression of noise by Fermi correlations 
in quantum point contacts under low temperatures, i.e., conductors with 
a small number of quantum modes.

On the other hand, when the ballistic transport is limited by a space charge,
Coulomb correlations may also result in a shot-noise suppression.
If the electron density injected into a ballistic conductor is low, 
the electron gas is nondegenerate, and Fermi statistical correlations
are not efficient.
For this case the Coulomb correlations are the main source of the shot-noise
suppression, as has been demonstrated by Monte Carlo simulations, 
\cite{gonzalez97,prb98} and subsequently analytically in a framework
of the Vlasov system of equations. \cite{prb00a}
In nanoscale devices, however, the injected carriers are usually degenerate,
which is due to a high level of contact doping and the elevated position 
of the Fermi level in the contact emitter. 
Therefore, it is of interest to consider the situation when both mechanisms, 
Fermi and Coulomb correlations, act together---the case that is important 
not only from a fundamental, but also from an applied point of view
\cite{likharev99} 
and has attracted less attention so far. \cite{naveh99,gonzalez99} 
In Ref.\ \onlinecite{naveh99} the problem for a multimode degenerate
conductor in the presence of a nearby gate has been posed, 
and the numerical results has been presented for a two-dimensional 
field-effect-transistor geometry.
Monte Carlo simulations in a two-terminal geometry, which take into account 
the degenerate injection from the contacts and Coulomb correlations 
in the ballistic region, have been performed. \cite{gonzalez99}
The relative significance of each mechanism in the shot-noise suppression 
and the limiting values for the noise suppression factors of each mechanism 
still remain unclear, since the analytical theory has not been proposed.

It is the objective of the present paper to address the problem 
of a shot-noise suppression under the conditions of the interplay between 
Fermi and Coulomb correlations in two-terminal multimode ballistic conductors.
To this purpose, we apply the recently developed analytical theory 
\cite{prb00b} for space-charge-limited (SCL) ballistic conductors 
to the case of a Fermi-Dirac degenerate injection.
Since we address the case of thick (in transversal dimensions) samples, 
the number of transversal modes (quantum channels) is large and
the dimensionality of a momentum space of electrons is three-dimensional (3D),
which makes a difference with the previous considerations of a one-channel 
or a few-channel quantum ballistic conductors (1D or quasi-1D momentum space).
\cite{blanter00,lesovik89,buttiker9092,land-martin9192}
Our analysis goes beyond the linear-response regime and zero-temperature
limit, -- the assumptions typically used to study a-few-channel conductors.
\cite{blanter00,lesovik89,buttiker9092,land-martin9192}
In a semiclassical framework, for a multimode ballistic conductor, 
we have derived analytical formulas that determine the {\em nonlinear} 
$I$-$V$ characteristics, the current-noise spectral density, and the shot-noise
suppression factors for each suppression mechanism in the limit of high biases.
We show that the Fermi shot-noise-suppression factor is limited below 
by the value determined by the properties of the injecting contact
(the ratio between the temperature and the Fermi energy), 
whereas the Coulomb noise suppression may be enhanced arbitrarily strong 
by extending the length of the ballistic sample with a simultaneous increase
of bias (provided the transport remains ballistic).
Therefore, the Coulomb suppression may be achieved much stronger even
in samples with a high degree of an electron degeneracy.

The paper is organized as follows.
In Sec.\ II we describe the semiconductor structure under consideration 
and discuss the main assumptions concerning the model.
In Sec.\ III we introduce the electron distribution function over 
the longitudinal injection energy, found by integrating over 
the transversal modes.
The analytical expression for the mean current is derived
as a function of the self-consistent potential barrier height.
Then, in the limit of high biases, the current-voltage characteristics 
beyond the Child approximation is obtained, which takes into account 
the degenerate Fermi-Dirac injection.
In Sec.\ IV the analytical expression for the suppressed value of the 
shot-noise power is derived,
in which the Fermi- and Coulomb-correlation contributions are distinguished.
The results for a particular GaAs semiconductor SCL diode are presented
in Sec.\ V. Finally, Sec.\ VI summarizes the main conclusions of the paper.

\section{The physical model}

We consider a two-terminal semiconductor ballistic sample with plane-parallel
heavily doped contacts at $x$=0 and $x$=$\ell$.
The structure may be considered as a $n$-$i$-$n$ SCL homodiode \cite{prb00a}
in which the current is determined by a charge injection from the contacts 
rather than by intrinsic carriers of the ballistic region. 
The applied bias $U$ between the contacts is assumed to be fixed by
a low-impedance external circuit and does not fluctuate.
In order to simplify the problem, we assume that due to the large difference 
in the carrier density between the contacts and the sample, and hence in 
the corresponding Debye screening lengths, all the band bending occurs 
in the ballistic base, and the relative position of the conduction band 
and the Fermi level does not change in the contacts. Therefore, 
when the bias is changed, the potential can vary exclusively inside 
the ballistic base, and the contacts are excluded from the consideration.
\cite{gonzalez97,prb98,prb00a}
The electron gas inside the contacts is assumed to be in thermal equilibrium.
However, in contrast to the previous works, \cite{gonzalez97,prb98,prb00a}
the Fermi level in respect to the bottom of the conduction band,
denoted here $\varepsilon_F$, may take not only negative, but positive values 
as well, i.e., the injected electrons may be either degenerate or 
nondegenerate, and follow, in general, the Fermi-Dirac distribution.
Assuming the transversal size of the conductor sufficiently thick
and high enough electron density, the electrostatic problem is considered 
in a one-dimensional plane geometry. \cite{prb00a} 

\section{Distribution function and mean current}

To describe the steady-state transport and low-frequency noise, we use
a semiclassical Vlasov system of equations, which consists of 
the collisionless Boltzmann transport equation for the distribution function 
and the Poisson equation for the self-consistent electrostatic potential. 
\cite{prb00a,prb00b}
Due to a stochastic nature of the injection, the distribution function 
and, consequently, the self-consistent potential both fluctuate in time.
The nonuniform distribution of the injected carriers leads to the creation of 
the potential minimum $\varphi_m$ at a position $x=x_m$.
The potential minimum acts as a barrier for the electrons by reflecting
a part of them back to the contact, thereby affecting the transport and noise 
properties.
It is the potential minimum fluctuations that induce the long-range
Coulomb interactions and lead to the suppression of the injected current 
fluctuations. \cite{prb00a}
We assume that the applied bias $qU > 5 k_BT$, where $q$ is the electron 
charge and $T$ is the temperature.
{}From this follows that the current is determined by only one injecting 
contact (at $x$=0 for definiteness), and the electrons from this contact 
that are able to pass over the barrier and arrive at the receiving contact 
at $x$=$\ell$ are all absorbed with the probability 1, 
since the corresponding energy states are empty.
All the electrons injected from the receiving contact are reflected back
because of the high-bias condition. Their contribution to the current
and noise is negligible.
Another assumption on the bias is $U_m \ll U < U_{cr}$, where 
$U_m\equiv -\varphi_m$ is the potential barrier height, and $U_{cr}$ is 
the bias at which the potential barrier vanishes. \cite{prb00a}
In this limit (``virtual-cathode approximation''), only the electrons 
that are able to pass over the fluctuating barrier contribute to the current 
and noise.
The nonhomogeneous electron density along the ballistic region 
is determined by \cite{prb00b} 

\begin{eqnarray} \label{den}
N(x)= \int_{\Phi_c}^{\infty} F_c(\varepsilon)
\frac{d\varepsilon}{2\sqrt{\varepsilon+\Phi(x)-\Phi_c}},
\end{eqnarray}
where $\Phi(x)=q\varphi(x)-q\varphi_m$ is the mean potential referenced 
to the minimum, with the value $\Phi_c\equiv\Phi(0)$ at the injecting 
contact.
It is clear that in such a definition the contact potential is equal to
the potential barrier height, $\Phi_c=qU_m$.
Note that $F_c(\varepsilon)$ is the distribution function over 
the longitudinal kinetic energy $\varepsilon$ at the injecting contact. 
Since during the ballistic motion only the longitudinal electron momentum
may vary, the injection distribution function is averaged over 
the transversal momentum ${\bf k}_{\perp}$:

\begin{equation} \label{fperp}
F_c(\varepsilon)= 2 \frac{\sqrt{2m}}{\hbar}
\int \frac{d {\bf k}_{\perp}}{(2\pi)^d} \, f(\varepsilon,{\bf k}_{\perp}),
\end{equation}
where $d$ is the dimension of a momentum space, $m$ the electron effective 
mass, $\hbar$ the Planck constant, $f(\varepsilon,{\bf k}_{\perp})$ 
the occupation number of a quantum state, the factor 2 takes into 
account the spin variable, and the additional factor $\sqrt{2m}/\hbar$ 
has been introduced for normalization convenience.
Assuming that the number of transversal modes is large, the dimension of 
a momentum space $d$=$3$, and we can perform integration over the transversal
states. Changing the variable of integration 
$d {\bf k}_{\perp}=(2\pi m/\hbar^2)d \varepsilon_{\perp}$,
where $\varepsilon_{\perp}$ is the transverse electron energy, and
taking into account that 
$f(\varepsilon,\varepsilon_{\perp})=f_F(\varepsilon+\varepsilon_{\perp})$,
with  $f_F(\varepsilon) = \{1+\exp[(\varepsilon -\varepsilon_F)/k_BT]\}^{-1}$
the Fermi-Dirac distribution, one gets

\begin{equation} \label{Fc}
F_c(\varepsilon)= \frac{N_c}{\sqrt{\pi k_B T}}
\ln \{1+\exp[(\varepsilon_F -\varepsilon)/k_BT]\},
\end{equation}
where $N_c=2(2\pi m k_B T)^{3/2}/(2\pi\hbar)^3$ is the effective density 
of states.
Integrating the distribution function (\ref{Fc}) over the energy,
one obtains the electron density injected from the contact,

\begin{eqnarray} \label{den0}
N_0 = \int_0^{\infty} F_c(\varepsilon)
\frac{d\varepsilon}{2\sqrt{\varepsilon}}
= \frac{N_c}{\sqrt{\pi}} \int_0^{\infty} \ln \,(1 + e^{\xi-z^2})\, dz,
\end{eqnarray}
where $\xi\equiv\varepsilon_F/k_BT$ is the reduced Fermi energy.
The injected electron density may also be expressed in a more familiar form

\begin{eqnarray} \label{N0}
N_0 =\frac{1}{2} N_c {\cal F}_{1/2}(\xi),
\end{eqnarray}
where ${\cal F}_{1/2}$ is the Fermi-Dirac integral of index $1/2$. 
Since the Fermi-Dirac integrals ${\cal F}_j$ of different indexes $j$ 
will be frequently used throughout the paper, their properties 
are summarized in the Appendix.
Note that $N_0$ is half of the contact electron density 
$N_c {\cal F}_{1/2}(\xi)$,
since only the electrons with positive momenta are injected into the sample.

Substituting the distribution (\ref{Fc}) into Eq.\ (\ref{den}) taken at 
$x$=$x_m$, one obtains the electron density at the potential minimum:

\begin{eqnarray} \label{Nm}
N_m = \int_{\Phi_c}^{\infty} F_c(\varepsilon)
\frac{d\varepsilon}{2\sqrt{\varepsilon-\Phi_c}}
= \frac{1}{2} N_c {\cal F}_{1/2}(\alpha),
\end{eqnarray}
where $\alpha=(\varepsilon_F-\Phi_c)/k_BT$ is the parameter characterizing 
the position of the Fermi energy with respect to the potential barrier.
The density $N_m$ is an important parameter for computing the
current noise, as will be seen below.

The steady-state current is obtained by \cite{prb00b}

\begin{equation} \label{I0}
I = \frac{qA}{\sqrt{2m}} \int_{\Phi_c}^{\infty} F_c(\varepsilon)\, 
d\varepsilon,
\end{equation}
where $A$ is the cross-sectional area.
Substituting the distribution function (\ref{Fc}), one gets

\begin{equation} \label{I}
I = I_F \int_0^{\infty} \ln \,(1 + e^{\alpha-y})\, dy
= I_F\, {\cal F}_1(\alpha),
\end{equation}
where $I_F=4\pi q A m (k_B T)^2/(2\pi\hbar)^3$ and ${\cal F}_1$ is
the Fermi-Dirac integral (see the Appendix).
It is seen that under the ballistic SCL conduction,
the current is determined by the relative position of the Fermi energy 
and the potential barrier through the parameter $\alpha$.
This is in contrast to the case of diffusive conductors,
in which the current is determined by scattering strength.
The parameter $\alpha$ summarizes the dependence of the current on 
the applied bias and the length of the conductor, since they both affect
the potential barrier height,
whereas the factor $I_F$ is independent of those characteristics.

\begin{figure}
\narrowtext
\epsfxsize=8.0cm
\epsfbox{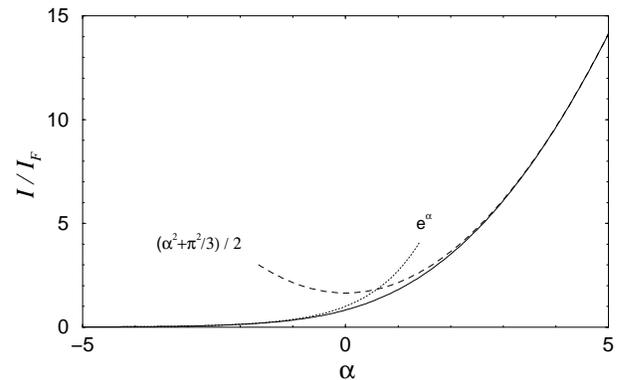}
\protect\vspace{0.5cm}
\caption{
Current as a function of the position $\alpha$ of the Fermi energy 
$\varepsilon_F$ in respect to the potential barrier $\Phi_c$, 
$\alpha=(\varepsilon_F-\Phi_c)/k_BT$.
The asymptotic approximations for nondegenerate and degenerate limits 
are plotted.}
\label{f1}
\end{figure}

Figure \ref{f1} illustrates the electric current as a function of $\alpha$
given by Eq.\ (\ref{I}). 
When the Fermi energy is sufficiently below the potential barrier, 
$\alpha<-3$, only the exponential tail of the contact distribution function 
is injected (nondegenerate injection limit).
Under this condition, according to the approximate formulas 
for the Fermi-Dirac integrals (\ref{FD1}), the current becomes

\begin{equation} \label{Iasym1}
I\approx I_F e^{\alpha}.
\end{equation}
When the Fermi energy is above the potential barrier by several $k_BT$,
it is the degenerate injection limit and, by using Eq.\ (\ref{FD2}), 
one gets the approximate formula for the current

\begin{equation} \label{Iasym2}
I \approx \frac{1}{2} I_F \left(\alpha^2+\frac{\pi^2}{3}\right).
\end{equation}
It is seen from Fig.\ \ref{f1}, that formula (\ref{Iasym2}) 
is accurate at $\alpha > 2$.
Note that the case of nondegenerate injection $\alpha < -3$, 
may occur when the contact electron density is either nondegenerate or
degenerate, depending on the position of the Fermi energy with respect to 
the conduction band edge characterized by the parameter $\xi$.
For $\xi<-3$, the contact electron density is nondegenerate, and 
this is the case of the Maxwell-Boltzmann injection, analyzed in detail in 
Ref.\ \onlinecite{prb00a}. 
Let us demonstrate that our formulas are in agreement with that case.
Equation (\ref{N0}) gives the injected electron density
$N_0 = \frac{1}{2} N_c \, e^{\xi}$, and
the current (\ref{Iasym1}) is expressed as

\begin{equation} \label{Iasym11}
I =I_{MB} \, e^{-U_m/k_BT}, 
\end{equation}
where $I_{MB}=I_F(2N_0/N_c)=qAN_0\sqrt{2k_BT/\pi m}$ is the emission
current for the Maxwell-Boltzmann distribution
[compare with Eq.\ (46) of Ref.\ \onlinecite{prb00a}].
For $\xi>-3$ and $\alpha<-3$, the injected electrons that pass over 
the barrier are nondegenerate, but the contact electrons are degenerate;
hence the approximate formula (\ref{Iasym11}) for the current is no longer 
valid, and one has to use a more general relation, Eq.\ (\ref{Iasym1}).

It should be also noted that in the general case of a Fermi-Dirac injection,
the contact emission current is $I_0=I_F{\cal F}_1(\xi)$.
This is the maximum (saturation) current that is achieved when the applied 
bias $U \ge U_{cr}$, the barrier vanishes ($\Phi_c=0$, $\alpha=\xi$), and
the conduction is no longer space-charge limited.
The current in units of its saturation value is simply

\begin{equation} \label{I2}
\frac{I}{I_0} = \frac{ {\cal F}_1(\alpha) }{ {\cal F}_1(\xi) }.
\end{equation}

It was shown in a previous paper \cite{prb00b} that the asymptotic 
behavior of the current in SCL ballistic conductors obeys the Child law 
in the leading-order terms independently of the injection distribution:

\begin{equation} \label{Ich}
I_{\rm Child} = \frac{4}{9}\kappa A \sqrt{\frac{2q}{m}}
\frac{U^{3/2}}{\ell^2},
\end{equation}
where $\kappa$ is the dielectric permittivity, and $\ell$ is the length
of the ballistic conductor.
However, this formula is only accurate at very high biases,
in the range where the SCL conduction is difficult to maintain. 
This is a consequence of a rough approximation, in which the velocity
spread of electrons at the potential minimum is neglected.
To obtain a satisfactory good approximation at lower biases, it is necessary 
to keep the next-order terms that are specific with respect to the injection 
distribution.
The general formula for an arbitrary injection function 
has been recently derived: \cite{prb00b}

\begin{equation} \label{Ich1}
I = I_{\rm Child}\, 
\left( 1+\frac{3}{\sqrt{qU}} 
\frac{\int_0^{\infty} F_c(\epsilon+\Phi_c) \epsilon^{1/2} d\epsilon}
{\int_0^{\infty} F_c(\epsilon+\Phi_c) d\epsilon} \right).
\end{equation}
In our case of the injection distribution function (\ref{Fc}), one finds 
the following expression 

\begin{equation} \label{Ich2}
I = I_{\rm Child}\, \left(
1 + \frac{3\sqrt{\pi}}{2} \sqrt{\frac{k_BT}{qU}}
\frac{{\cal F}_{3/2}(\alpha)}{{\cal F}_1(\alpha)} \right).
\end{equation}
In the nondegenerate limit, $\alpha<-3$, one obtains 
(${\cal F}_{3/2}/{\cal F}_1)\to 1$, and Eq.\ (\ref{Ich2}) leads to 
the Langmuir formula for the Maxwell-Boltzmann injection. \cite{prb00a} 
In the opposite limit of high degeneracy, $\alpha\gg 1$, one gets

\begin{equation} \label{Ichdeg}
I \approx I_{\rm Child}\, \left(
1 + \frac{8}{5} \sqrt{\frac{\varepsilon_F-qU_m}{qU}} \right),
\end{equation}
which can be used to estimate the current for Fermi ballistic conductors
beyond the Child approximation.
Here, we remark that in the degenerate limit and at $\varepsilon_F\gg qU_m$,
the current (\ref{Ichdeg}) is independent of temperature $T$ in both terms.
For an arbitrary degree of degeneracy, the general expression (\ref{Ich2})
can be used.

\section{Current noise}

To calculate the current noise, one has to define the partial injection 
current $I_c(\varepsilon)$ at the contact and its fluctuation properties. 
{}From Eq.\ (\ref{I}) it follows that 

\begin{equation} \label{Ic}
I_c(\varepsilon)=\frac{I_F}{k_BT} 
\ln [1 + e^{(\varepsilon_F-\varepsilon)/k_BT}],
\end{equation}
which corresponds to the current carried by electrons with injection 
(longitudinal) energies between $\varepsilon$ and $\varepsilon+d\varepsilon$, 
giving after the integration the total emission current 
$I_0=\int_0^{\infty} I_c(\varepsilon)d \varepsilon$.

\begin{figure}
\narrowtext
\epsfxsize=7.0cm
\epsfbox{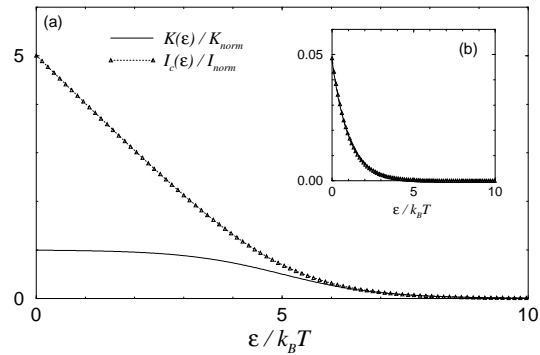}
\protect\vspace{0.5cm}
\caption{
Shot-noise power per unit energy $K(\varepsilon)$ and partial
current $I_c(\varepsilon)$ for injected electrons for two cases:
(a) degenerate, $\xi$=5; 
(b) nondegenerate, $\xi$=$-3$; 
Here, $K_{\rm norm}=2qI_F/k_BT$, $I_{\rm norm}=I_F/k_BT$.}
\label{f2}
\end{figure}

The correlation function for the fluctuations of the partial injection 
currents may be written generally as 

\begin{equation} \label{kuncor}
\langle\delta I_c(\varepsilon)\delta I_c(\varepsilon')\rangle = 
K(\varepsilon)(\Delta f)\delta(\varepsilon-\varepsilon'),
\end{equation}
where $\Delta f$ is the frequency bandwidth 
(we assume the low-frequency limit).
For the particular case of Fermi 3D injection, the function $K(\varepsilon)$
is determined by \cite{kogan,remark1}

\begin{equation} \label{K0}
K(\varepsilon) = 2q \frac{2qA}{\hbar} 
\int \frac{d {\bf k}_{\perp}}{(2\pi)^d} \, 
f(\varepsilon,{\bf k}_{\perp}) [1 - f(\varepsilon,{\bf k}_{\perp})].
\end{equation}
The integration over the transversal states may be performed explicitly
by taking into account that
$f_F(1-f_F)=-k_BT (\partial f_F/\partial \varepsilon)$ and
$\int_0^{\infty} d \varepsilon_{\perp} 
f_F(\varepsilon+\varepsilon_{\perp}) [1 - f_F(\varepsilon+\varepsilon_{\perp})]
=k_BT \, f_F(\varepsilon)$.
This gives a simple expression:

\begin{equation} \label{K}
K(\varepsilon) = \frac{2q I_F}{k_BT} f_F(\varepsilon),
\end{equation}
We remind the reader that $\varepsilon$ is the {\em longitudinal} energy 
component.
The Fermi factor $1-f_F$ has disappeared after the integration,
\cite{remark2} but the Fermi noise-suppression effect is present 
in Eq.\ (\ref{K}).
Indeed, in the degenerate limit $\varepsilon_F\gg k_BT$,
the partial current (\ref{Ic}) is a linear function of energy,
$I_c(\varepsilon)\sim (\varepsilon_F-\varepsilon)$, at $\varepsilon\to 0$
[see Fig.\ \ref{f2}(a)].
This occurs because of the increase of the number of transversal states
as the longitudinal energy $\varepsilon$ decreases.
Despite the increasing of the number of states, the shot-noise power
per unit energy represented by the function (\ref{K}) is constant 
at $\varepsilon\ll \varepsilon_F$ [Fig.\ \ref{f2}(a)].
As a result, $K(\varepsilon)/2qI_c(\varepsilon)\approx 1/\xi \ll 1$ 
at $\varepsilon\to 0$, indicating the noise suppression effect.
In contrast, for nondegenerate case, both functions 
$\propto \exp[(\varepsilon_F-\varepsilon)/k_BT]$, and 
$K(\varepsilon)/2qI_c(\varepsilon)\approx 1$ [Fig.\ \ref{f2}(b)],
which leads to the Poisson noise.
Additionally, we note that, since $I_F\sim T^2$, the injection noise vanishes,
$K(\varepsilon)\to 0$, in the limit $T\to 0$.

\begin{figure}
\narrowtext
\epsfxsize=8.0cm
\epsfbox{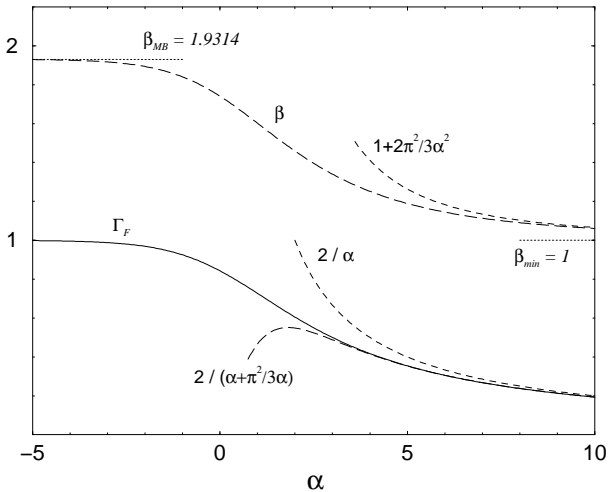}
\protect\vspace{0.5cm}
\caption{
Fermi shot-noise-suppression factor $\Gamma_F$ and shot-noise parameter 
$\beta$ as functions of the position of the Fermi energy $\alpha$. 
The asymptotic approximations for nondegenerate and degenerate limits 
are shown.}
\label{f3}
\end{figure}

The current-noise spectral density for the electron flow, when
Coulomb correlations are disregarded, is given by \cite{prb00b}

\begin{eqnarray} \label{Siuncor0}
S_I^{\rm uncor} = \int_{\Phi_c}^{\infty} K(\varepsilon) d\varepsilon.
\end{eqnarray}
Here, the integration is performed over the energies above the barrier
height $\Phi_c$, since only the electrons transmitted over the barrier 
contribute to the current noise at high biases.
Substitution of expression (\ref{K}) yields

\begin{eqnarray} \label{Siuncor}
S_I^{\rm uncor} = 2q I_F \ln \, (1 + e^{\alpha}) = 2q I_F {\cal F}_0(\alpha).
\end{eqnarray}
{}From this result we find the shot-noise-suppression factor caused
by Fermi correlations

\begin{equation} \label{supF}
\Gamma_F = \frac{S_I^{\rm uncor}}{2qI} = \frac{{\cal F}_0(\alpha)}
{{\cal F}_1(\alpha)}.
\end{equation}
This function is plotted in Fig.\ \ref{f3}.
It is clear that in the nondegenerate limit, one gets
$({\cal F}_0/{\cal F}_1)\to 1$, and obviously $\Gamma_F\to 1$.
An important feature is that $\Gamma_F$ is a decreasing function of $\alpha$.
In the degenerate limit, $\alpha\gg 1$, it approaches the asymptotic behavior:

\begin{equation} \label{supF2}
\Gamma_F \approx \frac{2}{\alpha+(\pi^2/3\alpha)}.
\end{equation}
It is seen from Fig.\ \ref{f3} that formula (\ref{supF2}) is accurate
at $\alpha>3$.
The limiting minimal value for $\Gamma_F$ occurs when the barrier
vanishes ($\alpha=\xi$),

\begin{equation} \label{supFmin}
\Gamma_F^{\rm min} = \frac{2}{\xi} = \frac{2k_BT}{\varepsilon_F}.
\end{equation}
The numerical factor in Eq.\ (\ref{supFmin}) depends on the dimensionality 
of a momentum space. By taking different values of $d$ in Eqs.\ (\ref{fperp})
and (\ref{K0}), one can get $\Gamma_F^{\rm min}$=$c\,k_BT/\varepsilon_F$ with
$c$=2 ($d$=3), $3/2$ ($d$=2), and 1 ($d$=1).
In all the cases, the shot-noise Fermi suppression is determined by the ratio 
between the temperature of the injected electrons $T$ and their Fermi energy 
$\varepsilon_F$. 
For a fixed $\varepsilon_F$, the suppression may be enhanced 
by decreasing the temperature $T\to 0$, but it is independent of the bias,
the ballistic length and the other parameters of the conductor.

The current noise, which takes into account both  Fermi and Coulomb 
correlations, is determined by \cite{prb00b}

\begin{equation} \label{Si0}
S_I = \int_{\Phi_c}^{\infty} \gamma^2(\varepsilon)\,
K(\varepsilon)\,d\varepsilon,
\end{equation}
where the energy-resolved shot-noise-suppression factor 

\begin{eqnarray} \label{gamma}
\gamma(\varepsilon) &=& \frac{3}{\sqrt{qU}}
[\sqrt{\varepsilon-\Phi_c} - \upsilon],
\end{eqnarray}
and the constant $\upsilon$ for an arbitrary injection distribution 
is given by \cite{prb00b}

\begin{equation} \label{upsilon0}
\upsilon = \frac{N_m}{F_c(\Phi_c)}
\end{equation}
By using Eqs.\ (\ref{Fc}) and (\ref{Nm}), we find for the Fermi 3D injection

\begin{eqnarray} \label{upsilon}
\upsilon = \frac{\sqrt{\pi k_BT}}{2} \, 
\frac{{\cal F}_{1/2}(\alpha)}{{\cal F}_0(\alpha)}.
\end{eqnarray}
Thus, for the current-noise power (\ref{Si0}), after using Eqs.\
(\ref{K}), (\ref{gamma}), and (\ref{upsilon}), we find 

\begin{equation} \label{Si}
S_I =  \beta \,2qI\, \frac{k_BT}{qU}.
\end{equation}
In this formula, the constant $\beta$ is determined only by $\alpha$:

\begin{eqnarray} \label{beta}
\beta(\alpha) = 9 \left(1 - \frac{\pi}{4}
\frac{[{\cal F}_{1/2}(\alpha)]^2}{{\cal F}_0(\alpha){\cal F}_1(\alpha)}\right),
\end{eqnarray}
To distinguish the noise suppression caused by different mechanisms,
one can define the shot-noise-suppression factor due to a pure Coulomb 
suppression

\begin{eqnarray} \label{supC}
\Gamma_C = \frac{S_I}{S_I^{\rm uncor}} = \beta \, \frac{k_BT}{qU} 
\frac{{\cal F}_1(\alpha)}{{\cal F}_0(\alpha)},
\end{eqnarray}
whereas the overall shot-noise-suppression factor becomes

\begin{equation} \label{sup}
\Gamma = \Gamma_C \Gamma_F = \frac{S_I}{2qI} =  \beta \, \frac{k_BT}{qU}.
\end{equation}
It is seen that the current noise may be suppressed by both
the temperature $T$ and the bias $U$. This is in contrast to 
the pure Fermi suppression (\ref{supFmin}), which is sensitive to $T$, 
but independent of the bias.
The dependence on $U$ comes from the Coulomb correlations
and originates from the function $\gamma(\varepsilon)$.
The coefficient $\beta$ is a parameter that depends on the degree
of degeneracy, as follows from Eq.\ (\ref{beta}).
For the Fermi 3D injection, $\beta$ is a decreasing function of $\alpha$
ranging between two limiting values (see Fig.\ \ref{f3}):
$\beta_{min} < \beta < \beta_{\rm MB}$, where 
$\beta_{\rm MB} = 9 (1-\pi/4)\approx 1.9314$
is a limiting value in the nondegenerate limit (Maxwell-Boltzmann injection), 
and $\beta_{min}$=1 is a limiting value in the degenerate limit.
For high degeneracy, the approximate formula for $\beta$ may be obtained
by using the expansions for the Fermi-Dirac integrals (\ref{FD2}),
one gets 

\begin{eqnarray} \label{beta2}
\beta \approx 1 + \frac{2}{3}\frac{\pi^2}{\alpha^2}, \qquad \alpha\gg 1.
\end{eqnarray}
Figure \ref{f3} demonstrates the validity of such an approximation.
The asymptotic behavior of the Coulomb suppression factor $\Gamma_C$ in a 
high degenerate limit is obtained as

\begin{eqnarray} \label{supuncor2}
\Gamma_C \approx \frac{1}{2}\frac{\varepsilon_F-\Phi_c}{qU},
\quad \varepsilon_F-\Phi_c \gg k_BT,
\end{eqnarray}
which takes the minimal value at $U=U_{cr}$:

\begin{eqnarray} \label{supuncormin}
\Gamma_C^{\rm min} \approx \frac{\varepsilon_F}{2qU_{cr}}.
\end{eqnarray}
It should be emphasized that the difference between the two noise-suppression 
mechanisms is fundamental: While $\Gamma_F$ cannot be decreased 
further by varying the parameters of the conductor, since its minimal value 
is fixed by the contact properties 
[by the parameter $\xi$ as follows from Eq.\ (\ref{supFmin})].
In contrast, the factor $\Gamma_C$ may be decreased by increasing 
the ballistic length $\ell$ of the conductor, since for longer conductors
the critical bias $U_{cr}$ under which the barrier disappears is higher,
and $\Gamma_C$ may drop deeper. 
As a consequence, $\Gamma$ may also attain much lower values.
It is important to highlight, that in both nondegenerate and degenerate 
limits, the total shot-noise suppression factor $\propto k_BT$, 
and can therefore be reduced by decreasing the temperature.

\section{Example}

To illustrate the results, consider the GaAs $n$-$i$-$n$ ballistic diode 
of length $\ell$=0.5$\,\mu$m at $T$=4 K.
For this temperature and $m$=0.067$m_0$, the effective density of states is
$N_c\approx 6.7\times 10^{14}$ ${\rm cm}^{-3}$.
Assuming the contact doping $1.6\times 10^{16}\,{\rm cm}^{-3}$, the reduced
Fermi energy $\xi \approx 10$, and the contact electrons are degenerate.
For this set of parameters, the Debye screening length associated with
the contact degenerate electron density is approximately
$L_D=\sqrt{\kappa k_BT/[q^2 N_c {\cal F}_{-1/2}(\xi)]}\approx 14$ nm.
Since $L_D\ll\ell$, the space-charge effects and, therefore, 
the Coulomb shot-noise suppression are important in a wide range of biases.

\begin{figure}
\narrowtext
\epsfxsize=8.0cm
\epsfbox{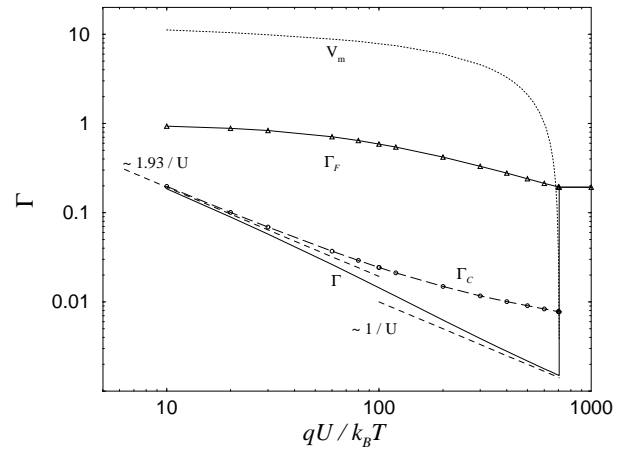}
\protect\vspace{0.5cm}
\caption{
Shot-noise-suppression factors: $\Gamma_F$ (Fermi), $\Gamma_C$ (Coulomb), and 
$\Gamma$=$\Gamma_F\Gamma_C$ (total), and potential barrier height 
$V_m$ as functions of applied bias $U$. 
The asymptotic limiting lines $\beta_{\rm MB}(k_BT/qU)$ and $k_BT/qU$ 
for $\Gamma$ are shown by dashes.}
\label{f4}
\end{figure}

Let us introduce the normalized biases: $V=qU/k_BT$ and $V_m=qU_m/k_BT$.
The calculation of the steady-state potential profile for different biases $V$
shows that the potential barrier varies from $V_m\approx 11.2$
at $V$=10 to $V_m$=0 at $V$=$V_{cr}\approx 705$ ($U_{cr}\approx 243$ mV) 
(see Fig.\ \ref{f4}).
In this range, the charge-limited conduction is controlled by the barrier
height $V_m$, and by increasing the bias, one can observe the crossover from
nondegenerate ($\alpha=\xi-V_m<-1$) to degenerate ($2<\alpha<10$) injection.
This crossover is illustrated in Fig.\ \ref{f4}, where the shot-noise 
suppression factors $\Gamma$, $\Gamma_F$, and $\Gamma_C$ are plotted
as functions of bias.
Indeed, the Fermi suppression factor $\Gamma_F$ varies from 1 at low biases 
to $2/\xi\approx 0.2$ at high biases, in agreement with formulas 
(\ref{supF})--(\ref{supFmin}).
Moreover, the factor $\Gamma$ lies between two asymptotic lines:
$\beta_{\rm MB}(k_BT/qU)$ at low biases (nondegenerate limit)
and  $k_BT/qU$ at high biases (degenerate limit), in agreement 
with Eq.\ (\ref{sup}) and the variation of $\beta$ in Fig.\ \ref{f3}.
The Coulomb correlation factor $\Gamma_C$ decreases with bias up to
the lowest value $\approx 0.0078$ at $U$=$U_{cr}$. After that value it 
increases sharply to 1 due to the disappearance of the potential barrier. 
The sharp increase of $\Gamma_C$ at $U=U_{cr}$ is discontinuous in this
asymptotic theory, which neglects the high-order terms in the expansions.
The exact calculations \cite{unpub} give a smoother behavior.
Note that both mechanisms essentially suppress shot noise at large $U$,
but $\Gamma_C$ is always much lower than $\Gamma_F$ under SCL conditions.

\section{Summary}

In conclusion, we have derived the analytical formulas that describe
the mean current and the shot-noise power in degenerate space-charge-limited 
ballistic conductors.
In the framework of a semiclassical Vlasov system of equations, which takes 
into account the fluctuations of the potential profile self-consistently, 
we have obtained a deep shot-noise suppression of more than two orders of
magnitude caused by two independent mechanisms: Fermi and Coulomb correlations.
The derived formulas clearly distinguish the shot-noise suppression factors 
caused separately by Fermi correlations (\ref{supF}), Coulomb correlations 
(\ref{supC}), and by the joint action of both [Eq.\ (\ref{sup})].

We show that the Fermi shot-noise-suppression factor is limited below by the 
ratio between the temperature and Fermi energy of the contact electrons.
The Coulomb noise-suppression factor, however, may attain much lower values
$\varepsilon_F/2qU$, because of its dependence on the applied bias 
$U\gg k_BT/q$.
The asymptotic behavior of the overall shot-noise-suppression factor
in a high degenerate limit
was found to be $k_BT/qU$, independently of the material parameters.
Finally, for the degenerate Fermi-Dirac injection, the asymptotic formula 
for the mean current beyond the Child approximation is proposed.
\cite{remark3}

\acknowledgements

We are grateful to V.\ A.\ Kochelap for valuable discussions.
This work has been partially supported by the Generalitat de Catalunya,
Spain, and the NATO linkage Grant No.\ HTECH.LG 974610.

\appendix
\section{Fermi-Dirac functions and their approximations}

The Fermi-Dirac functions are encountered, whenever one wants
to describe the electronic transport in degenerate semiconductor 
or metallic systems, and they are defined as \cite{blakemore}

\begin{equation} \label{FD}
{\cal F}_j(\alpha) = \frac{1}{\Gamma(j+1)} 
\int_0^{\infty} \frac{y^j\, dy}{1+e^{y-\alpha}},
\end{equation}
where $\Gamma(j)$ is the gamma function of the index $j$.
For the expressions of this paper, the $\Gamma$ functions take the values
$\Gamma(\frac{3}{2})=\sqrt{\pi}/2$, $\Gamma(\frac{5}{2})=3\sqrt{\pi}/4$,
$\Gamma(1)=\Gamma(2)=1$.

For positive indexes $j$, the Fermi-Dirac integrals
can also be rewritten [obtained by integrating by parts (\ref{FD})]:

\begin{equation} \label{FD0}
{\cal F}_j(\alpha) = \frac{j}{\Gamma(j+1)} 
\int_0^{\infty} y^{j-1} \, \ln \,(1+e^{\alpha-y})\, dy,
\end{equation}

A simple relation between the integrals of different orders is

\begin{equation}
{\cal F}_j(\alpha) = d{\cal F}_{j+1}/d\alpha.
\end{equation}

Unfortunately, these integrals cannot be resolved analytically except for 
the trivial cases:

\begin{equation}
\matrix{
{\cal F}_0(\alpha) = \ln(1+e^\alpha),         & & j=0  \cr
{\cal F}_{-1}(\alpha) = (1+e^{-\alpha})^{-1}, & & j=-1. }
\end{equation}

However, for small and large $\alpha$, one may use the approximate formulas
\cite{blakemore} for the {\em nondegenerate limit, $\alpha < -2$},

\begin{equation} \label{FD1}
{\cal F}_j(\alpha) = e^{\alpha} \qquad \forall \, j,
\end{equation}

and for the {\em degenerate limit, $\alpha\gg 1$},

\begin{equation} \label{FD2}
{\cal F}_j(\alpha) = \frac{\alpha^{j+1}}{(j+1)\,\Gamma(j+1)} 
\left[ 1+ \frac{\pi^2}{6} \frac{j(j+1)}{\alpha^2} + 
O\left(\frac{1}{\alpha^4}\right) \right].
\end{equation}

\end{multicols}

\begin{references}
\bibitem{remark2}
The disappearance of the factor $1-f_F$ in Eq.\ (\ref{K}) leads to 
a remarkable difference between the multichannel 3D case and a one-channel
case studied earlier (Ref.\ \onlinecite{land-martin9192}).
In a 1D conductor the noise comes from a narrow energy range of order $k_BT$ 
at the Fermi level, while in a 3D conductor the main contribution to the noise
comes from the lowest longitudinal energies far below the Fermi energy,
where more transversal states within the range $k_BT$ near the Fermi sphere 
are available [see also discussion of Fig.\ \ref{f2}(a)].
\bibitem{remark3}
Note that in all the asymptotic formulas, for both the current and noise,
the bias $U$ may be replaced by the more exact $U+U_m$.
Because of our assumption $U_m\ll U$, we neglect $U_m$.
This correction may improve, however, the results at intermediate biases
when the barrier is not sufficiently small.
\end{references}
\end{document}